\documentclass[preprint,12pt]{elsarticle}

\usepackage{amsmath}
\usepackage{color}

\begin{document}

%\preprint{}

\title{Magnetic phase diagram, magnetotransport and inverse magnetocaloric effect in the noncollinear antiferromagnet Mn$_5$Si$_3$}



%\author{}
%\thanks{}
%\affiliation{}


\author{Roberto F. Luccas}
\address{Instituto de Ciencia de Materiales de Madrid, Consejo Superior de Investigaciones Cient\'{\i}ficas (ICMM-CSIC), Sor Juana In\'es de la Cruz 3, 28049 Madrid, Spain.}
\address{Instituto de F\'{\i}sica Rosario, CONICET-UNR, Bv. 27 de Febrero 210bis, S2000EZP Rosario, Santa F\'e, Argentina.}
\author{Gabriel S\'anchez-Santolino}
\address{Instituto de Ciencia de Materiales de Madrid, Consejo Superior de Investigaciones Cient\'{\i}ficas (ICMM-CSIC), Sor Juana In\'es de la Cruz 3, 28049 Madrid, Spain.}
\author{Alex Correa-Orellana}
\address{Instituto de Ciencia de Materiales de Madrid, Consejo Superior de Investigaciones Cient\'{\i}ficas (ICMM-CSIC), Sor Juana In\'es de la Cruz 3, 28049 Madrid, Spain.}
\address{Laboratorio de Bajas Temperaturas y Altos Campos Magn\'eticos, Unidad Asociada UAM, CSIC, Departamento de F\'isica de la Materia Condensada, Instituto de Ciencia de Materiales Nicol\'as Cabrera and Condensed Matter Physics Center (IFIMAC), Universidad Aut\'onoma de Madrid, Spain}
\author{Federico J. Mompean}
\address{Instituto de Ciencia de Materiales de Madrid, Consejo Superior de Investigaciones Cient\'{\i}ficas (ICMM-CSIC), Sor Juana In\'es de la Cruz 3, 28049 Madrid, Spain.}
\author{Mar Garc\'{\i}a-Hern\'andez}
\address{Instituto de Ciencia de Materiales de Madrid, Consejo Superior de Investigaciones Cient\'{\i}ficas (ICMM-CSIC), Sor Juana In\'es de la Cruz 3, 28049 Madrid, Spain.}
\author{Hermann Suderow}
\address{Laboratorio de Bajas Temperaturas y Altos Campos Magn\'eticos, Unidad Asociada UAM, CSIC, Departamento de F\'isica de la Materia Condensada, Instituto de Ciencia de Materiales Nicol\'as Cabrera and Condensed Matter Physics Center (IFIMAC), Universidad Aut\'onoma de Madrid, Spain}

\date{\today}

\begin{abstract}
The antiferromagnet Mn$_5$Si$_3$ has recently attracted attention because a noncollinear spin arrangement has been shown to produce a topological anomalous Hall effect and an inverse magnetocaloric effect. Here we synthesize single crystals of  Mn$_5$Si$_3$ using flux growth. We determine the phase diagram through magnetization and measure the magnetoresistance and the Hall effect. We find the collinear and noncollinear antiferromagnetic phases at low temperatures and, in addition, a third magnetic phase, in between the two antiferromagnetic phases which has ferromagnetic character. The latter magnetic phase might be caused by strain produced by Cu inclusions that lead to quenched fluctuations of the mixed character magnetic ordering in this compound.
\end{abstract}

\maketitle

\section{Introduction}

Topologically non-trivial magnetic structures, such as skyrmions in magnets, can be used to develop new devices for storage or computation.\cite{nagaosa13}
Skyrmions were observed in the compound MnSi which has helical magnetic ordering.\cite{muhlbauer09,du14,neubauer09,thessieu97,lebech95,demishev12,pappas09,okada01,petrova09,koyama00,jeong04}
Other compounds in the Mn-Si family remain however poorly studied. Attention has been given to the compound Mn$_5$Si$_3$, due to the appearance of a topological anomalous Hall effect in thin films showing an antiferromagnetic (AFM) phase with non-collinear magnetic order and of an inverse magnetocaloric effect.\cite{vinokurova90,surgers14,chen14,surgers16,surgers17,songlin02,biniskos18}

Mn$_5$Si$_3$ crystallizes in a hexagonal structure (Pearson Symbol $hP16$ space group $P6_3/mcm$).\cite{villarsbook} The Mn ions have two sites in the crystallographic structure, Mn1 and Mn2\cite{surgers14,surgers16,surgers17,biniskos18}. A magnetic transition is observed at $T_{N2}$ $\approx$ 95 K\cite{vinokurova90,gottschilch12,silva02,brown95,lander67,alkanani95,biniskos18}, from paramagnetic to AFM, which goes together with a change in the crystal structure. The atomic positions in one of the Mn sites, Mn2, are divided into two sets\cite{biniskos18}. Magnetic order occurs in the undivided site Mn1 and in two thirds of the Mn atoms of the other site Mn2. When lowering the temperature below $T_{N1}$ $\approx$ 65 K, the crystal structure of Mn$_5$Si$_3$ loses inversion symmetry and magnetic moments reorient to form a noncollinear phase\cite{biniskos18}. The magnetic arrangement is quite complex, being both noncollinear and noncoplanar. As in the phase at higher temperatures, atoms of the Mn1 site carry a magnetic moment together with two thirds of the atoms of the other site Mn2. This distribution of magnetic moments remains the same as in the high temperature phase, but the magnitude of magnetic moments at the Mn2 sites depends on their position in the crystal lattice\cite{gottschilch12,silva02,brown95,lander67,alkanani95,biniskos18}.

In systems with noncoplanar spin arrangements, a finite Berry phase induces a ficticious magnetic field which can produce the so called topological Hall effect. \cite{lee07,nagaosa10,ye99,bruno04} A comparative study of thin films of antiferromagnetic Mn$_5$Si$_3$ and of C-doped ferromagnetic Mn$_5$Si$_3$ shows that the topological Hall effect develops in the low temperature noncollinear AFM phase.\cite{surgers14,surgers16,surgers17,brown92,silva02}

One the other hand, a strong inverse magnetocaloric effect has been found and associated to the same noncollinear AFM phase. The magnetocaloric effect stems from the Maxwell relation between magnetic field induced changes in the entropy and temperature induced changes in the magnetization, $\left( \frac{\partial S}{\partial B}\right)_T = \left( \frac{\partial M}{\partial T}\right)_B$. In a simple spin system, the changes in the magnetization versus temperature (decrease with decreasing temperature) can be used to cool with by adiabatically demagnetizing a previously isothermally magnetized sample. In systems showing magnetic order, the functional dependence of the magnetization with temperature strongly depends on the magnetic field. This can lead to situations where magnetic field induced entropy variations are such that cooling is obtained by adiabatic magnetization, instead of adiabatic demagnetization. This is called the inverse magnetocaloric effect\cite{FRANCO2018112}. The inverse magnetocaloric effect has been observed close to first order ferromagnetic phase transitions and close to some AFM transitions\cite{Liu2012}. The inverse magnetocaloric effect was observed in polycrystals of Mn$_5$Si$_3$ and associated to the collinear to noncollinear AFM transition\cite{songlin02,biniskos18}.

Here we study magnetization and magnetotransport in Mn$_5$Si$_3$. We observe the two mentioned magnetic AFM transitions. The low temperature noncollinear phase is shifted to lower temperatures with respect to previous observations, down to $T_{N1}^*$ $\approx$ 45 K.  We observe an additional intermediate phase between $T_{N1}$ $\approx$ 65 K and  $T_{N1}^*$.  We build the magnetic field-temperature phase diagram and study the magnetoresistance and the Hall effect, from which we obtain that the intermediate phase is likely a weak ferromagnetic phase. We find that the sign change of the magnetic entropy value leads to an inverse magnetocaloric effect below $T_{N1}^*$.

\section{Experimental}

We grew crystals of Mn$_5$Si$_3$ out of Cu flux. In solution with Cu, the melting temperature of Mn-Si mixtures decreases below 1000 $^\circ$C and there are no known ternary compounds between Cu, Mn and Si.\cite{canfieldbook,canfield92,okamoto91,chakraborti89,olesinski86}
Solution with Cu also helps avoiding the high vapor pressure of Mn at the melting point of Mn$_5$Si$_3$ (near 1300 $^\circ$C), which would produce Mn losses during growth. We filled Alumina crucibles with 3/13 at.\% Mn, 3/13 at.\% Si and 7/13 at.\% Cu.
We thus worked with a slight Si excess to compensate for the low Si mobility on Cu-Mn-Si phase diagram.\cite{johosonbook}
The excess is low enough to avoid growth of MnSi, which requires a ratio of 1:1.96 in excess of silicon when using Cu flux.\cite{canfieldbook,canfield92,johosonbook}
We used Mn from GoodFellow 99.95\%, Si from Alfa Aesar 99.9999\% and Cu from Aldrich Chemical Company Inc. 99.999+\%.
We sealed the Alumina crucibles in a controlled He atmosphere in quartz ampoules.
We heated to 1190 $^\circ$C in 3 hours, remained at this temperature for 3 hours and then cooled down to 900 $^\circ$C in 250 hours.
We then spinned the ampoules in a centrifuge to separate crystals from flux.\cite{canfieldbook,canfield92}
We obtained needle shaped crystals with 4 mm to 8 mm in length and an octogonal cross section of 1 mm$^2$ with white silver-like shiny surfaces.

X-Ray diffraction experiments were performed using a High Resolution X-ray Bruker AXS D8 ADVANCE Diffractometer at room temperature with Cu-K$\alpha$1 and Cu-K$\alpha$2 radiation.
Rietveld annalysis was performed on X-Ray diffractograms using FullProf tool.
For the scanning transmission electron microscope characterization, we used an aberration-corrected JEOL JEM-ARM 200cF electron microscope equipped with a cold field emission gun and a Gatan Quantum electron energy-loss spectrometer.
For local spectrum imaging, an electron energy-loss spectrum was acquired for every pixel when scanning the beam with an acquisition time of 0.25 s per pixel.
Four wire resistivity measurements were performed between 1.8 K and 300 K applying a current of 10 mA.
Magnetization experiments were performed in a Quantum Design SQUID magnetometer.
Experiments of magnetization as a function of temperature (M {\emph{vs.}} T) as well as magnetic field (M {\emph{vs.}} H) were performed from 0.05 T to 7 T and in the temperature range from 1.8 K to 300 K.
Magnetic field was applied parallel to the long axis of the needles (c-axis of the hexagonal crystal structure).
In order to maximize the magnetic signal, we measured a group of several needles with same orientation (60.74 mg of mass).
Results coming from M {\emph{vs.}} T and the temperature derivative of $M T$ {\emph{vs.}} T, show peaks and features that we associate to magnetic transition Temperatures;\cite{fisher62} we use them to build the Magnetic Field-Temperature phase diagram. We used a Quantum Design PPMS system to make magnetoresistance and Hall effect measurements. The sample was positioned perpendicular to the magnetic field, so that the magnetic field was oriented in plane. Current was always applied parallel to the long axis of the needles, or the c-axis of the crystal structure.

\section{Results}

\begin{figure}
\begin{center}
\includegraphics[width=0.6 \columnwidth]{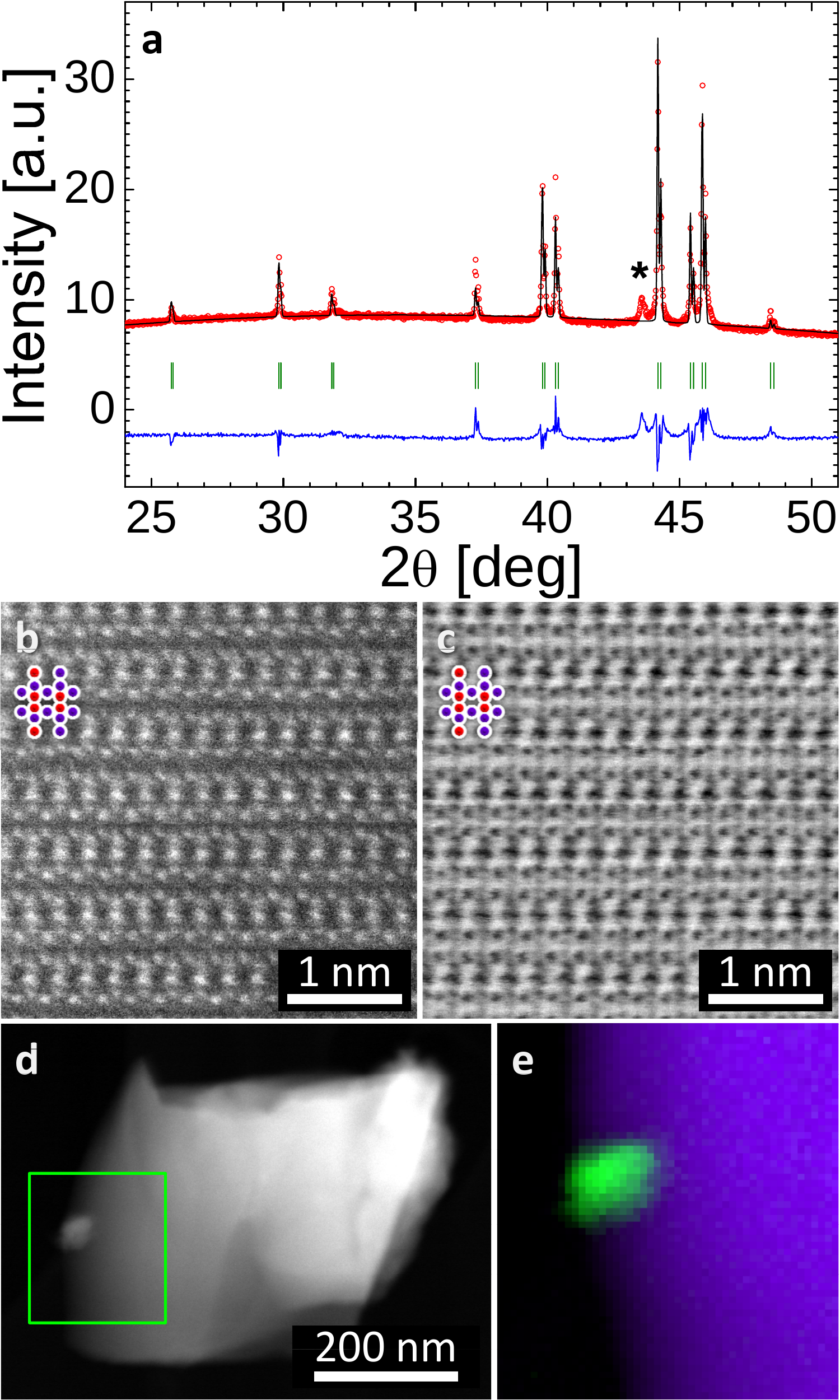}
\vspace{-10.5pt}
\caption{\label{Fig1}{\bf a}
X-Ray diffractogram of Mn$_5$Si$_3$ powder.
Experimental data are shown as red dots, expected peak's positions as green vertical lines and the correspoding fit to the diffractogram as a black line.
The difference between measurement and fit is shown as a blue line.
The black star highlights the only peak observed that is not Mn$_5$Si$_3$, it can be adscribed to pure Cu.
{\bf b} Scanning Transmission Electron Microscopy (STEM), High Angle Annular Dark Field (HAADF) and
{\bf c}
Annular Bright Field (ABF) images of a Mn$_5$Si$_3$ single crystal along the (100) direction.
We superimpose the unit-cell projection, marking the Mn (purple) and Si (red) atomic positions.
{\bf d}
Low magnification HAADF image of particles found attached to the surface of Mn$_5$Si$_3$ crystals.
{\bf e}
Electron energy-loss spectroscopy (EELS) image providing a compositional map of the area marked by a green square in {\bf d}.
The overlaid integrated intensity signals coming from Mn $L_{2,3}$ are purple color and Cu $L_{2,3}$ edges are green color.}
\end{center}
\end{figure}

\subsection{Crystal structure and composition of Mn$_5$Si$_3$}

Fig.\,\ref{Fig1}(a) shows the obtained diffractogram for X-ray diffraction experiments performed on Mn$_5$Si$_3$ powder ground out of our single crystals.
We find excellent matching with the expected series of peaks from Mn$_5$Si$_3$,\cite{aronsson60,amark36} with no traces of another intermetallic compound of the Mn-Si phase diagram.
A Rietveld analysis perform to X-Ray diffractogram points out values for crystal structure parameters in a = 6.913(2) {\AA} and c = 4.821(2) {\AA}, in good agreement with previous studies.\cite{aronsson60,amark36,leciejewicz08}
An additional phase, $\beta$-Mn$_5$Si$_3$ with W$_5$Si$_3$ structure (Pearson Symbol $tI32$ space group $I4/mcm$) and different values of crystal structure parameters, has been observed in nanowires of Mn$_5$Si$_3$.\cite{higgins11}
We found no trace of such a phase in our single crystals.
On the other hand, we found small amounts of Cu flux, corresponding to less than 3\% in volume.
An extra peak at 2$\theta$ = 43.6$^\circ$ is observed (black star) and corresponds to that Cu contamination coming from Cu flux.
To analyze the microstructure of the crystals, we acquired atomic resolution images by Scanning Transmission Electron Microscopy (STEM), High Angle Annular Dark Field (HAADF) and Annular Bright Field (ABF).
Both, HAADF (Fig.\,\ref{Fig1}(b)) and ABF (\ref{Fig1}(c)) show consistently a continous and nicely formed crystalline lattice, with the expected Mn$_5$Si$_3$ atomic arrangements.
In particular, ABF is sensitive to light elements and shows clearly the Si columns.\cite{findlay10}
For instance, we superimpose the Mn$_5$Si$_3$ atomic structure with red and purple dots marking Si and Mn positions respectively.

We also found small particles of approximately 50 nm in size attached to crystalline Mn$_5$Si$_3$ bulk (Fig.\,\ref{Fig1}(d)).
In order to determine the composition of those particles, we performed local electron energy-loss spectrum (EELS) analysis.
The color map in Fig.\,\ref{Fig1}(e) shows the overlaid integrated intensity signals for the Mn $L_{-2,3}$ edge (purple) and the Cu $L_{2,3}$  edge (green).
Clearly, attached particles are mostly composed of Cu, which is not present inside crystalline grains. So that the Cu flux present at samples is separated from the pure Mn$_5$Si$_3$ crystalline phase.

\subsection{Magnetic phase diagram}

\begin{figure}
\includegraphics[width=\columnwidth]{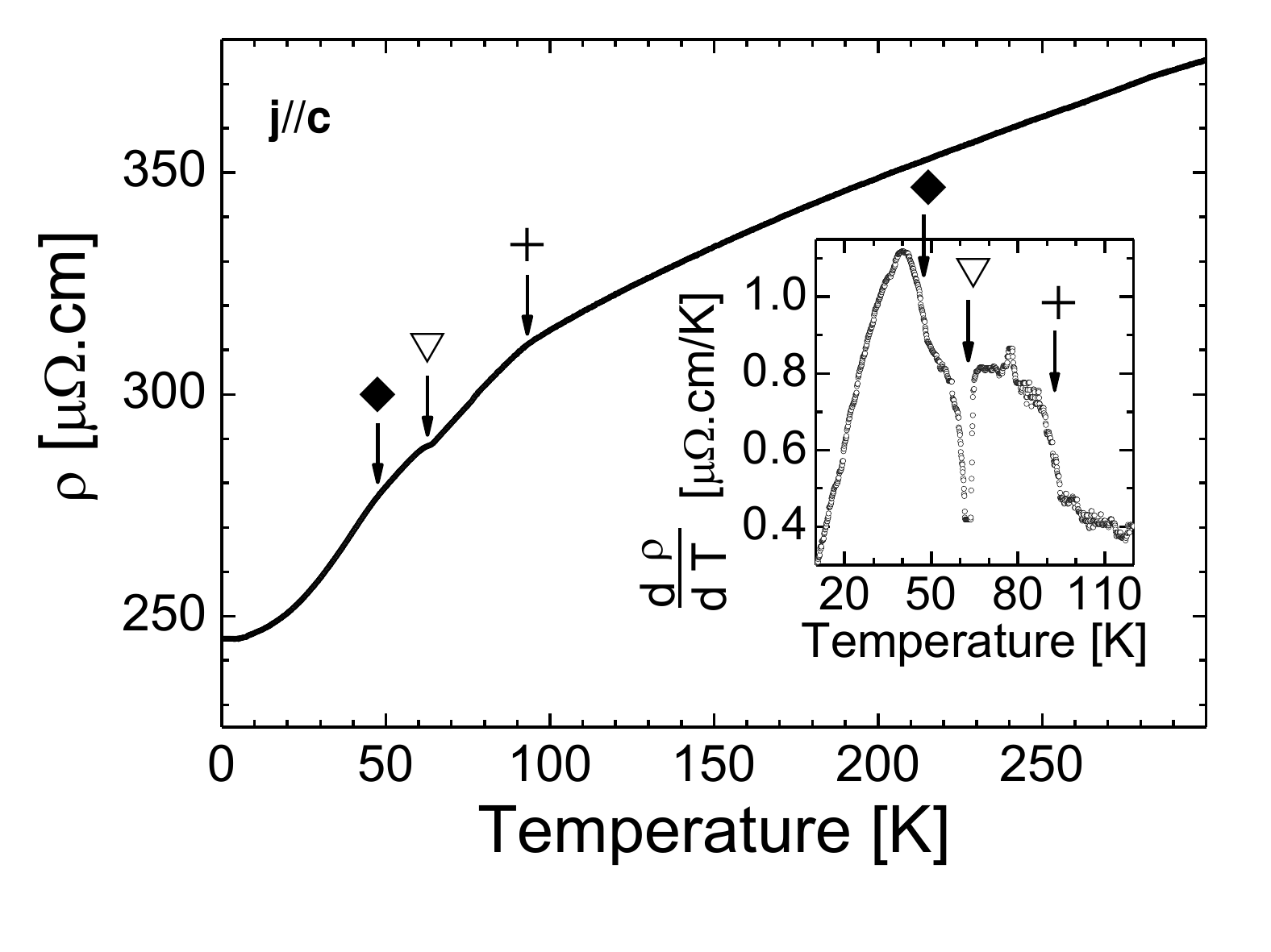}
\vspace{0pt}
\caption{\label{Fig2}
Resistivity $\rho$ {\emph{vs.}} Temperature T in single crystalline Mn$_5$Si$_3$ at zero magnetic field.
Features related to $T_{N2}$ at approximately 95 K (plus symbol), $T_{N1}$ at approximately 65 K (open triangle) and $T_{N1}^*$ at approximately 45 K (filled diamond) are highlighted by arrows.
Inset shows the Temperature derivative of $\rho$ as a function of T.}
\end{figure}

In Fig.\,\ref{Fig2} we show resistivity $\rho$ as a function of temperature T at zero magnetic field.
The inset shows the slope of $\rho(T)$, $\frac{d\rho}{dT}$, which we use to identify the different transition temperatures and establish the phase diagram.
Between about 100 K and room temperature we observe an approximately linear temperature dependence.
Below 100 K we observe several kinks and features marked by different symbols in Fig.\,\ref{Fig2}.
We observe step-like change in $\frac{d\rho}{dT}$ at $T_{N2}$ = 95 K (plus symbol), a small plateau at $T_{N1}$ = 65 K (open triangle) and a new step-like change at $T_{N1}^*$ = 45 K (filled diamond).
At low temperatures, the resistivity continues dropping, with a non-linear behavior as a function of T.
The resistivity decreases by about 35\% between room temperature and low temperatures; similarly as in thin films, with a decrease of about 15\%.\cite{surgers14}
This indicates that magnetic scattering limits the electronic conduction in this material. Measurements in single crystals show two sharp features at $T_{N2}$ and $T_{N1}$ that are much better resolved than in thin films\cite{surgers16}. We do not observe such sharp features here.

\begin{figure}
\begin{center}
\includegraphics[width=0.7 \columnwidth]{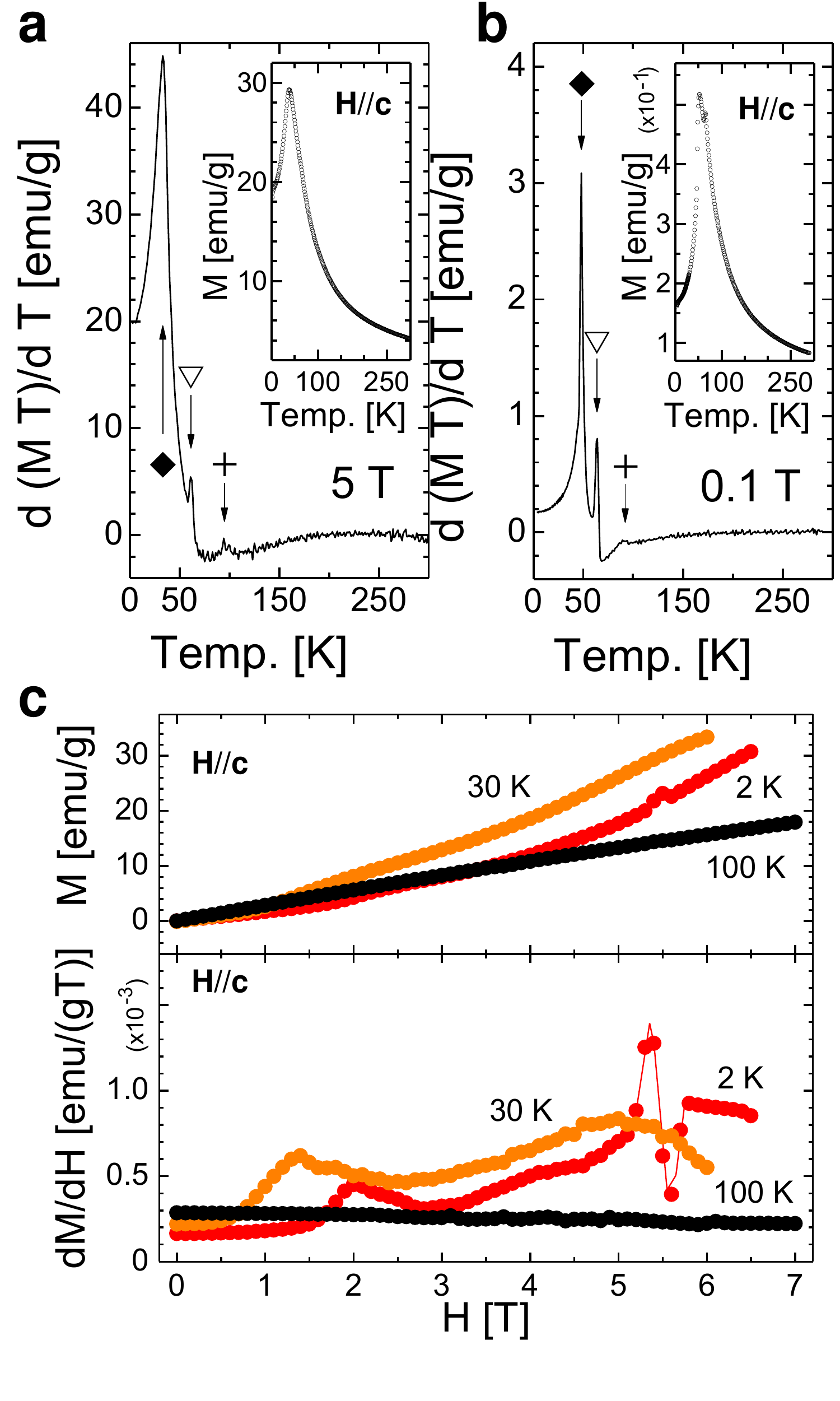}
\vspace{-21,96063pt}
\caption{\label{Fig3} 
Normalized magnetic moment M as a function of Temperature T and magnetic field H, and their corresponding derivatives, for selected data values.
In {\bf a} we show the temperature derivative of $MT$ as a function of temperature for 5 T and in {\bf b} for 0.1 T. Insets show corresponding M {\emph{vs.}} T original data sets. Transitions are marked as $T_{N1}^*$, filled diamonds, $T_{N1}$, open triangles and $T_{N2}$, plus symbols. In {\bf c} we show M {\emph{vs.}} H and is magnetic field derivative for a representative set of temperatures, below $T_{N1}^*$ at 2 K (red) and  at 30 K (orange) and in the paramagnetic phase at 100 K (black).}
\end{center}
\end{figure}

In Fig.\,\ref{Fig3}(a,b) we show the temperature derivative of $MT$ as a function of T for 5 T and 0.1 T (insets show $M(T)$). We highlight $T_{N2}$, $T_{N1}$ and $T_{N1}^*$ with crosses, open triangles and filled diamonds respectively.
We observe changes in the transition temperatures induced by the magnetic field below $T_{N1}^*$. In Fig.\,\ref{Fig3}(c) we show magnetization M as a function of magnetic field H (top) as well as their magnetic field derivative (bottom) for temperatures well above $T_{N2}$ and below $T_{N1}^*$. We observe a double peak between 2 T and 3 T at low temperatures, indicating that the phase below $T_{N1}^*$ changes when applying a magnetic field.

\begin{figure}
\begin{center}
\includegraphics[width=0.9 \columnwidth]{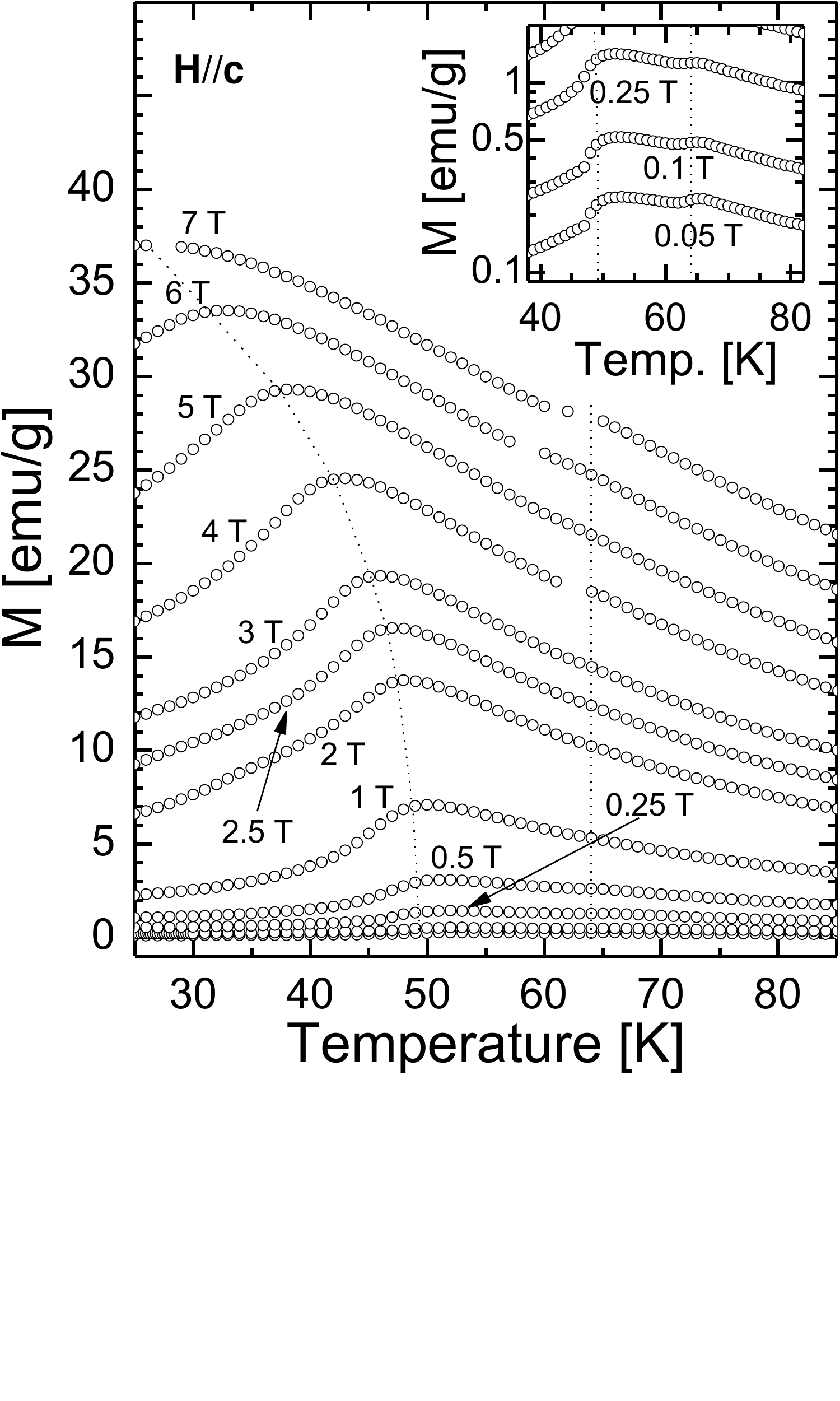}
\vspace{-106,22047pt}
\caption{\label{Fig4} Magnetization as a function of temperature for different values of magnetic field applied (from 0.05 T to 7 T). Inset shows data at the lowest magnetic fields (from 0.05 T to 0.25 T).}
\end{center}
\end{figure}

In Fig.\,\ref{Fig4} we show the magnetization as a function of temperature between 0.05 T and 7 T. We focus on $T_{N1}$ and $T_{N1}^*$, observed at 65 K and 45 K
respectively at zero magnetic field. In the inset we show a detail of the results at the lowest magnetic fields, from 0.05 T to 0.25 T. In this range, we clearly see two features that evolve differently with the magnetic field. These features can be further identified in the main panel. The feature at $T_{N1} = 65 K$ remains independent of the magnetic field, whereas $T_{N1}^*$ moves to lower temperatures with magnetic field, reaching 30 K at 7 T.

\begin{figure}
\includegraphics[width=\columnwidth]{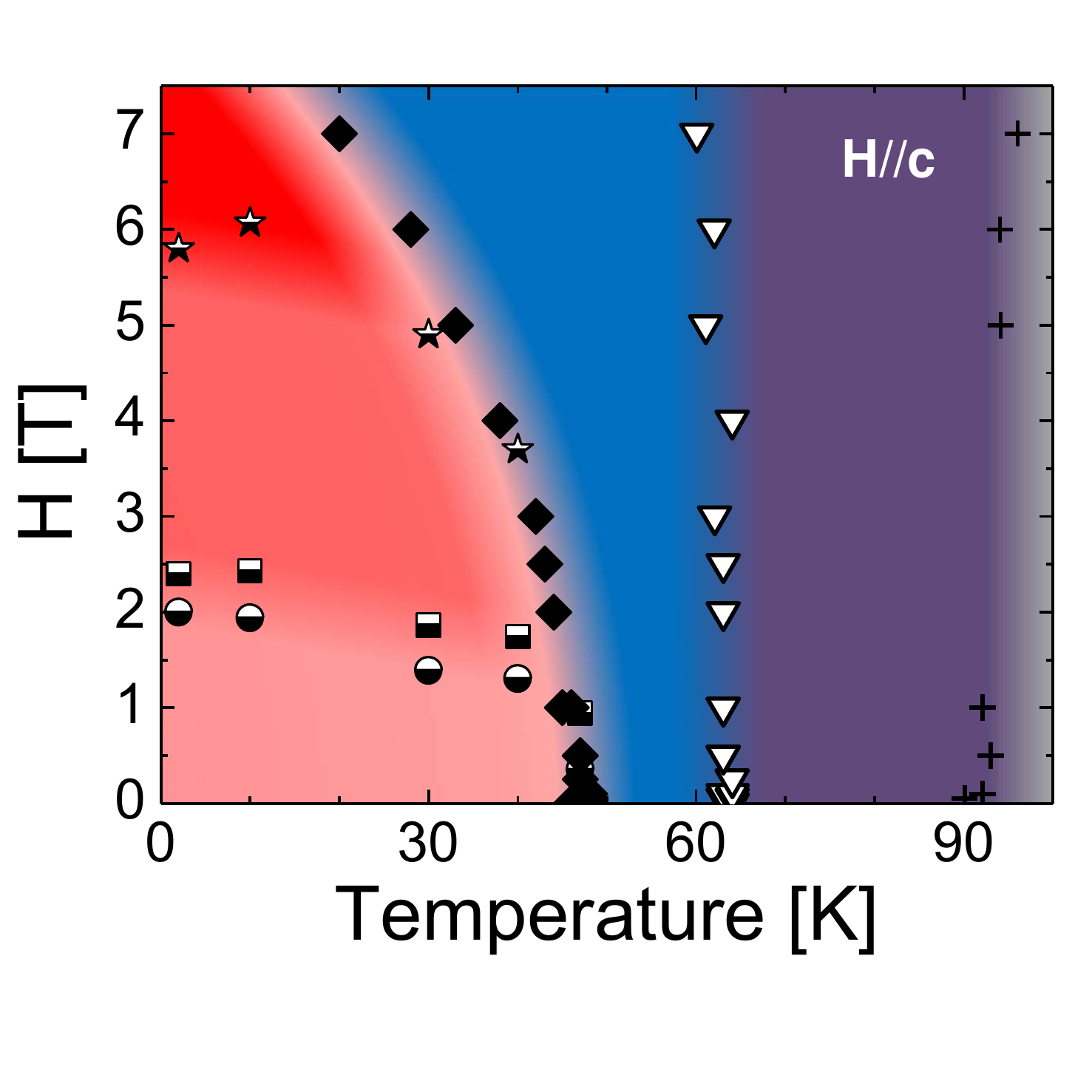}
\vspace{0 pt}
\caption{\label{Fig5}
The magnetic phase diagram as a function of temperature (H {\emph{vs.}} T) of our flux grown single crystals of Mn$_5$Si$_3$. We observe three transitions as a function of T.
The two at higher temperatures, $T_{N2}$ (plus symbols) and $T_{N1}$ (open triangles), are independent of the magnetic field.
The third one, $T_{N1}^*$ (filled diamonds) is strongly dependent on H.
Partially filled symbols (squares, circles and stars) correspond to features observed on measurements of the magnetization as a function of the magnetic field.
They are asociated to metamagnetic transitions in the phase below $T_{N1}^*$. Filled symbols and plus provide data obtained from measurements as a function of temperature.}
\end{figure}

With these results, we build the magnetic phase diagram shown in Fig.\,\ref{Fig5}.
The low temperature phase has several transitions with magnetic field and is strongly field dependent. However, the two high temperature phases remain practically insensitive to the magnetic field.

\subsection{Magnetoresistance and Hall effect}

\begin{figure*}
\includegraphics[width=\textwidth]{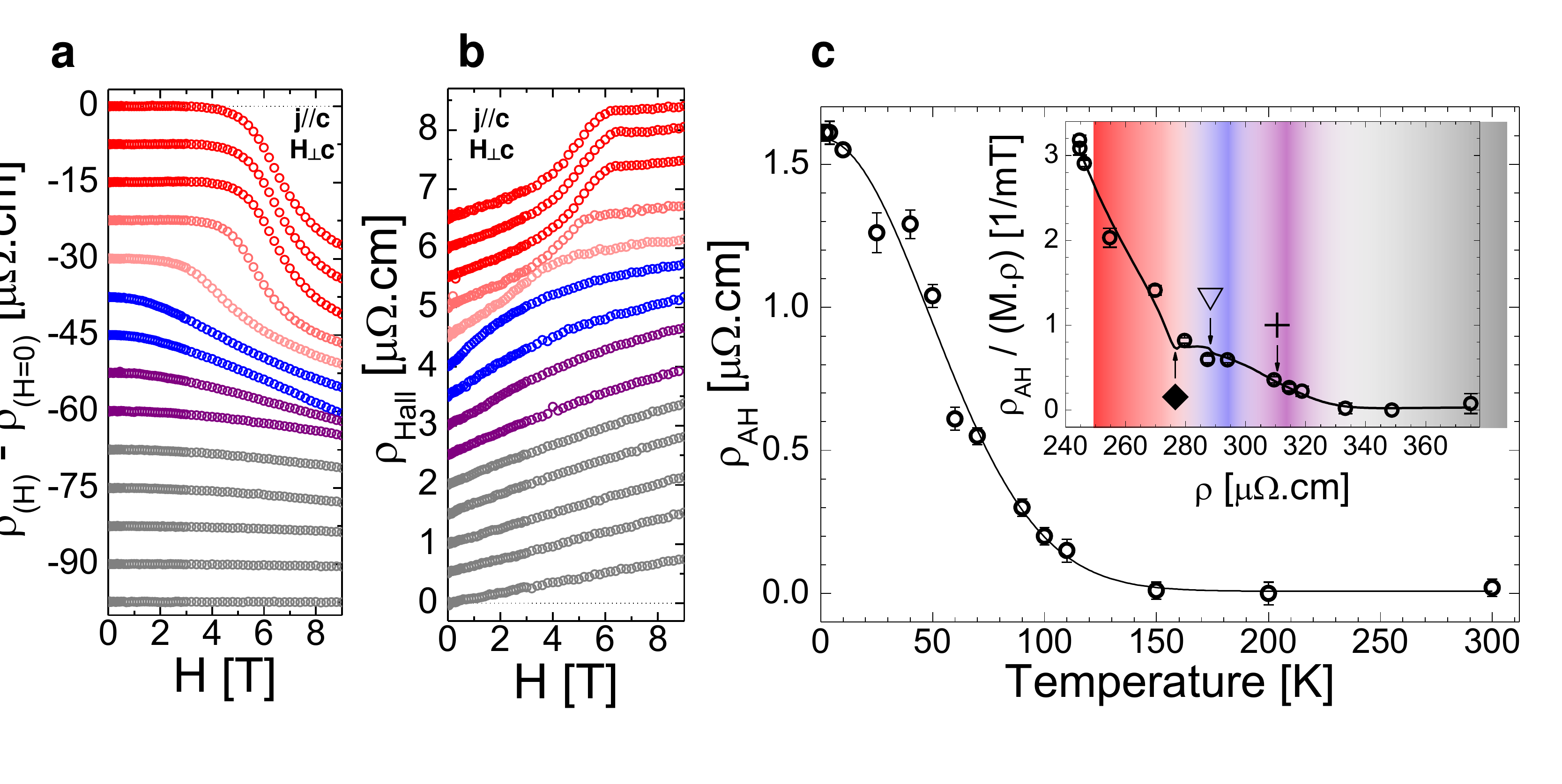}
\vspace{0pt}
\caption{\label{Fig6}{\bf a} Magnetoresistance $\rho(H)$ and {\bf b}
Hall resistivity $\rho_{Hall}$ of Mn$_5$Si$_3$ as a function of magnetic field H for different temperatures (from top to bottom, 2 K, 4 K, 10 K, 25 K, 40 K (in red), 50 K, 60 K (in blue), 70 K, 90 K (in purple) and 100 K, 110 K, 150 K, 200 K and 300 K (in grey)).
Curves are shifted by a fixed amount. The colors represent the different phases, as discussed in Figure 5. {\bf c} Anomalous contribution to the Hall resistivity $\rho_{AH}$ as a function of temperature. We extrapolate the high magnetic field part of $\rho_{Hall}$ shown on in {\bf b} to obtain $\rho_{AH}$. The line is guide to the eye and is obtained by a polynomial fitting of the points. In the inset we show $\rho_{AH}$ divided by the magnetization M and the resistivity $\rho$ as a function of the resistivity $\rho$ as open circles. The line shows the same quantity obtained using the line that serves as a guide in the main panel. In the inset, we have added a background to show the temperature ranges of each phase, as in Figure 5. The different magnetic transitions are marked by plus symbols, open triangle and filled diamond (for, respectively $T_{N2}$, $T_{N1}$ and $T_{N1}^*$).}
\end{figure*}

In Fig.\,\ref{Fig6}(a) we show the magnetoresistance $\rho(H)$ - $\rho(H=0)$ versus the magnetic field H. At high temperatures in the paramagnetic phase there is no appreciable magnetoresistance. However, at lower temperatures the magnetoresistance starts showing strong differences. First, in all cases, the magnetoresistance is negative. Then, below $T_{N2}$ (purple curves in Fig.\,\ref{Fig6}(a)), the magnetoresistance has a decrease with magnetic field with a parabolic behavior. Between $T_{N2}$ and $T_{N1}$, the magnetoresistance is nearly linear. Below $T_{N1}^*$, we observe a well defined plateau which evolves when drecreasing temperature. At the lowest temperatures, the plateau remains up to 4 T. At high magnetic fields, the magnetoresistance strongly decreases, with a tendency to show saturation at the highest magnetic fields.

In Fig.\,\ref{Fig6}(b) we show the Hall coefficient $\rho_{Hall}$ as a function of magnetic field H. Above $T_{N2}$, in the paramagnetic phase, we observe a Hall coefficient proportional to the magnetic field. The slope $\frac{d\rho_{Hall}}{dH}$ at low fields increases with decreasing temperature. Between $T_{N2}$ and $T_{N1}$ $\frac{d\rho_{Hall}}{dH}$ is more pronounced than in the paramagnetic phase. Between $T_{N1}$ and $T_{N1}^*$ there is a change in the magnetic field dependence, with a strong $\frac{d\rho_{Hall}}{dH}$ at low fields and a saturation of $\rho_{Hall}$ at higher fields. Below $T_{N1}^*$, $\frac{d\rho_{Hall}}{dH}$ decreases again for low magnetic fields. In particular, in the field range of the plateau in the magnetoresistance, $\frac{d\rho_{Hall}}{dH}$ remains roughly at a fixed value. But at higher magnetic fields, coincinding with the drop in the magnetoresistance, the Hall effect increases strongly and becomes again proportional to the magnetic field at the highest fields.

\section{Discussion}

The magnetic phase diagram of Mn$_5$Si$_3$, shown in Fig.\,\ref{Fig5}, presents four phases at zero field, the non-magnetic high temperature phase, and three magnetic phases. N\'eel temperatures of the high temperature phases, $T_{N1}$ and $T_{N2}$, are independent on magnetic field. $T_{N1}^*$ depends on the magnetic field and the phase appearing below  $T_{N1}$ remains to lower temperatures. There are additional magnetic field induced phases at low temperatures below $T_{N1}^*$.

The Hall effect $\rho_{Hall}$ shows a linear magnetic field dependence in the paramagnetic phase at high temperatures, as expected for a normal metal. The increase of the $\frac{d\rho_{Hall}}{dH}$ when decreasing temperature in the paramagnetic phase can be related to a slight change in the carrier density, which we estimate to be from 1.04$\times 10^{22}$ cm$^{-3}$ to 0.56$\times 10^{22}$ cm$^{-3}$ between room temperature and $T_{N2}$. For lower temperatures we observe clearly an anomalous Hall effect (AHE).

The AHE consists on a finite zero field extrapolation of $\rho_{Hall}$ to zero magnetic field and indicates that there is magnetic skew scattering, in addition to the usual Lorentz force induced electron motion.\cite{nagaosa10}
The AHE appears together with magnetic order and is much more pronounced at the low temperature phase below $T_{N1}^*$, where the high field plateau yields a large zero field extrapolation. The AHE corresponds to a finite Hall voltage even in absence of external magnetic field and requires a finite magnetic polarization together with a skewing mechanism for electron orbits, such as spin orbit coupling or magnetic impurities. Alternatively, electron trajectories might have a finite curvature due to a non-trivial topological feature in the magnetic order, which produces the topological Hall effect.

We can write $\rho_{Hall}$ as a sum of two terms: an ordinary one and an anomalous one. Empirically, it has been observed that\cite{zeng06}

\begin{equation}\label{eqone}
\rho_{Hall} = \rho_{OH} + \rho_{AH} = R_0 B + R_s 4 \pi M \qquad
\end{equation}

\noindent where $\rho_{OH}$ is the ordinary Hall resistivity in a perpendicular magnetic field $B$, $\rho_{AH}$ the anomalous Hall resistivity, and $R_0$ and $R_s$ the ordinary and anomalous Hall coefficients respectively.

The $\rho_{AH}$ term can be extracted from high magnetic field values, extrapolating $\rho_{Hall}$ linearly to zero.\cite{zeng06,lee07}
Circles in Fig.\,\ref{Fig6}(c) show $\rho_{AH}$ as a function of temperature obtained from $\rho_{Hall}$. For the extrapolation, we consider only magnetic fields above 7 T.
$\rho_{AH}$ is practically zero at high temperatures and increases when decreasing temperature.

To determine the different contributions to the Hall coefficient, we can follow the analysis of Refs.\,\cite{zeng06,surgers14} and write:

\begin{equation}\label{eqtwo}
\rho_{AH} = a \rho + b \rho^2 \qquad 
\end{equation}

\noindent where $\rho$ is the longitudinal magnetoresistance asociated to $\rho_{Hall}$.
The two contributions can be associated to skew scattering ($a$) \cite{smith55} and to a combination of an intrinsic Berry-phase curvature correction and an extrinsic phase winding from skew scattering at impurities ($b$).\cite{sundaram99,berger70} 
Due to the temperature variation of all involved quantitities ($R_s$, $M$, $\rho$, $a$ and $b$), it is always complex to solve eq. \ref{eqtwo} and separate all the effects, particularly the two contributions to $b$. We can assume that skew scattering is usually linear in magnetization $M$ ($a/M = constant$). With this, we can obtain $a$ and $b$ from eq. \ref{eqtwo} by plotting $\rho_{AH}/ (M \rho)$ as a function of $\rho$.\cite{zeng06}

Assuming that the temperature variation of $\rho_{AH}$ has a smooth behavior,\cite{zeng06} we fit $\rho_{AH}$ to an arbitrary polynomial (black line on Fig.\,\ref{Fig6}(c)). We then divide the fit by $(M \rho)$, obtaining $\rho_{AH}/ (M \rho)$. We plot $\rho_{AH}/ (M \rho)$ as a function of $\rho$ in the inset of Fig.\,\ref{Fig6}(c). The contribution to the AHE occurs essentially in the magnetic phases, developing below $T_{N2}$. Eventually, short range order close to the transition in the paramagnetic phase provides for a slight increase in $\rho_{AH}$  when approaching $T_{N2}$ from above.
We can identify two diferent kinds of behavior. For the AFM phases below $T_{N1}^*$ and between $T_{N1}$ and $T_{N2}$, we observe roughly a linear variation of $\rho_{AH}/ (M \rho)$ with $\rho$, which implies $b / M = constant$. For the intermediate phase between $T_{N1}^*$ and $T_{N1}$ we observe a plateau in $\rho_{AH}/ (M \rho)$ with $\rho$.

The different behaviour observed for $\rho_{AH}/(M  \rho)$ in the high and low temperature AFM phases implies that the ratio $a/M$ has a constant value that changes from phase to phase. It is much larger in the noncollinear low temperature AFM phase. As for $b / M$, we find $b / M \approx -0.017 (mT \mu \Omega cm)^{-1}$ between $T_{N1}$ and $T_{N2}$, and approximately four times this value, $b / M \approx -0.071 (mT \mu \Omega cm)^{-1}$ in the low temperature phase below $T_{N1}^*$. But in between $T_{N1}$ and $T_{N1}^*$ $b$ is zero.

\begin{figure}
\includegraphics[width=\columnwidth]{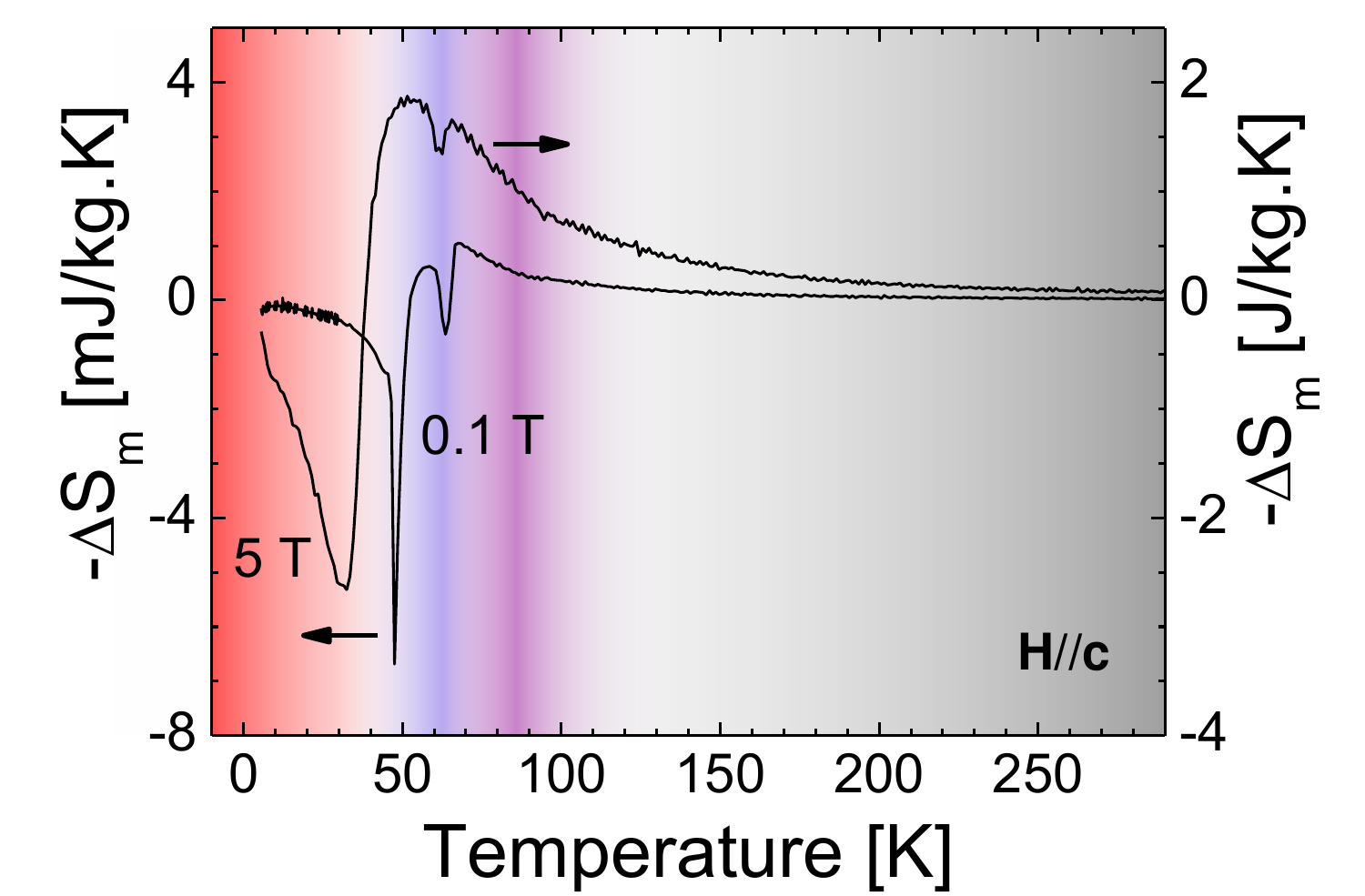}
\vspace{0pt}
\caption{\label{Fig7}
The magnetic entropy $\Delta S_m=S(T,B)-S(T,0)=\int_0^B \left( \frac{\partial M}{\partial T}\right)_B dB$ as a function of Temperature T for magnetic fields between 0 and 0.1 T and for magnetic fields between 0 and 5 T. The background color provides the different magnetic phase, as described in Figure 5. We observe an inverse magnetocaloric effect (negative $-\Delta S_m$) below $T_{N1}^*$.}
\end{figure}

An AHE with $b = 0$ implies that in the phase between $T_{N1}$ and $T_{N1}^*$ is dominated by usual magnetic skew scattering. This has been observed until now only in pure ferromagnets at low temperatures.\cite{fert72} Thus, the phase between $T_{N1}$ and $T_{N1}^*$ should have ferromagnetic character.

Our results for the low and high temperature AFM phases is in good agreement with neutron scattering experiments.\cite{biniskos18} Neutron scattering mostly focus on the behavior at finite q, without showing clear features between $T_{N1}$ and $T_{N1}^*$. Furthermore, the intensity at the AFM wavevectors at 50 K (between $T_{N1}$ and $T_{N1}^*$) is measured as a function of the magnetic field. Contrary to what we observe (our phase diagram, Fig.\,5), neutron scattering finds two magnetic transitions at 3 T and at 7 T, similar to those we find in the phase below $T_{N1}^*$. This points out that the transition to the noncollinear phase has been shifted down to $T_{N1}^*$ in our experiments, giving way to a new intermediate phase. 

Using our measurements, we can calculate the magnetic entropy variation $\Delta S_m=S(T,B)-S(T,0)=\int_0^B \left( \frac{\partial M}{\partial T}\right)_B dB$. Fig. \,\ref{Fig7} shows $-\Delta S_m$ in our Mn$_5$Si$_3$ samples, for variations of 0 T - 0.1 T and 0 T - 5 T. The inverse magnetocaloric effect occurs for negative $-\Delta S_m$ and we indeed find it in the low temperature phase below $T_{N1}^*$. For all other phases, the inverse magnetocaloric effect is absent. The value of $-\Delta S_m$ is remarkably large, as reported in this and other compounds with noncollinear AFM order\cite{songlin02,biniskos18}. The transition between the intermediate and high temperature magnetic phases (between $T_{N1}$ and  $T_{N2}$) also gives a spike, but $-\Delta S_m$ does not become negative.

Clearly, there is a game with the local magnetic moments that enters into the entropy and produces a situation which increases spin disorder with the magnetic field, an effect which is a priori counterintuitive. This can occur in Mn$_5$Si$_3$ where there are many sites without magnetic moment and the magnetic order is distributed among different Mn sites. Results of Ref.\,\cite{biniskos18} show that the magnetic excitation spectrum of the high temperature AFM phase is unusual, with coexisting spin waves of the AFM phase and spin fluctuations\cite{songlin02}. The increase in spin disorder has been associated to a magnetic field induced restoration of the high temperature collinear AFM phase and an increase in fluctuations.

In our samples, we find that the transition to the noncollinear state is shifted to lower temperatures, with an additional magnetic phase between the collinear and noncollinear AFM orders. The additional magnetic phase does not influence the low temperature non-collinear phase, which has all features observed previously, as the anomalous Hall effect and the inverse magnetocaloric effect. The additional phase might be due to  magnetic fluctuations quenched by some mechanism.

Mn$_5$Si$_3$ is likely sensitive to strain or stress, which might either favor or quench fluctuations. Our crystals are mostly small and needle like, with clear facets and cross sections with polygon shape. Previous work was made in thin films\cite{surgers14,surgers16}, in single crystals grown out of a stoichiometric melt\cite{surgers17}, in polycrystals \cite{songlin02} and in large, tens of grams weight, Czrochalski grown single crystals\cite{biniskos18}. The role of stress in all these samples is not easy to understand by now. In principle, flux growth provides the smallest amount of stress, because samples grow in the liquid. However, it is quite clear that the Cu inclusions in our samples might lead to stress that quenches fluctuations and produces the additional phase and thus play a relevant role.

Following the doping or pressure dependence of the resistivity, or studying the elastoresistivity could lead to new insight. In particular, it would be interesting to check the influence of modifying the composition on the inverse magnetocaloric effect. When exchanging Mn by Fe, the inverse magnetocaloric effect disappears, although there are strong magnetic entropy (positive $-\Delta S_m$) changes occuring at higher temperatures as the Mn moments are forced into a ferromagnetic arrangement.\cite{songlin02,Candini20046819} To obtain a negative  $-\Delta S_m$ at higher temperatures, we would need to increase the temperature for the appearance of noncollinear magnetic order,\cite{FRANCO2018112,gschneidnerjr05} shifting at the same time the fluctuating magnetic phase to higher temperatures. 

Previous measurements of the magnetoresistance in single crystals provide beautiful and sharp features and a magnetic phase diagram that closely resembles the one obtained in neutron scattering\cite{surgers16,surgers17}. The anomalous Hall effect is observed, as well as the magnetic field induced transitions in the low temperature noncollinear AFM phase. There are strong hysteresis effects in the magnetoresistance and in the Hall effect, when the field is applied along an axis within the plane\cite{surgers16,surgers17}. Furthermore, switching effects are observed in the noncollinear phase. There are transitions between different magnetic field induced changes in the spin arrangement, which are seen as sign changes in the Hall resistivity but not in the magnetization. In our Hall effect measurements, we apply the current along the c-axis and the magnetic field within the plane, a configuration where the switching effects are small.\cite{surgers17} On the other hand, the resistivity at zero magnetic field shows sharp features and the zero temperature extrapolation is sensibly below our value in the single crystals of Refs.\,\cite{surgers16,surgers17}. This indicates that our samples have a larger degree of disorder or suffer stress due to the Cu inclusions.

\section{Summary and conclusions}

In summary, we have grown crystals of Mn$_5$Si$_3$ using flux of Cu and determined the magnetic phase diagram of this compound. We observe a transition from a paramagnetic to an AFM phase at $T_{N2}$ and a transition from AFM to another magnetic phase at $T_{N1}$. From an analysis of the magnetoresistance and the Hall effect we obtain that this phase is possibly of weak ferromagnetic character. Below $T_{N1}^*$ spins reorganize into a non collinear AFM phase. There is a strong inverse magnetocaloric effect associated with the low temperature non-collinear AFM phase.

%\appendix
%\section{}

\section{Acknowledgments}
This work was supported by the Spanish MINECO (Consolider Ingenio Molecular Nanoscience CSD2007-00010 program, FIS2017-84330-R, MDM-2014-0377, MAT2014-52405-C2-2-R, FJCI-2015-25427 and CSD2009-00013), by the Comunidad de Madrid through program NANOMAGCOST-CM (S2018/NMT-4321) and MAD2D-CM (S2013/MIT-3007) and by EU (Graphene Core1 contract No. 696656, Nanopyme FP7-NMP-2012\_SMALL-6 NMP3-SL-2012\_310516 and COST CA16218).
We acknowledge I. Guillam\'on for guidance in preparing the magnetotransport experiment and handling the small needles as well as for discussions and M. Varela and J. M. Gonz\'alez-Calbet for help with the electron microscope.
Electron microscopy obsevations carried out at the Centro Nacional de Microscopia Electr\'onica-UCM.
We acknowledge technical support by SEGAINVEX at UAM.
We acknowledge C. Munuera and C. E. Sobrero for fruitful discussions.

\bibliographystyle{elsarticle-num-names}
\bibliography{RFLuccas-etal-References}

\end{document}